\RequirePackage{fix-cm}
\documentclass[smallextended]{svjour3}       % onecolumn (second format)
\smartqed  % flush right qed marks, e.g. at end of proof
\usepackage{graphicx}
\usepackage{color}

\newcommand{\be}{\begin{equation}}
\newcommand{\ee}{\end{equation}}
\newcommand{\bea}{\begin{eqnarray}}
\newcommand{\eea}{\end{eqnarray}}
\newcommand{\bdm}{\begin{displaymath}}
\newcommand{\edm}{\end{displaymath}}
\newcommand{\lb}{\label}

\begin{document}

\title{Phantom singularities and their
quantum fate: general relativity and beyond
}
\subtitle{A CANTATA COST Action topic}

\titlerunning{Phantom singularities}        % if too long for running head

\author{Mariam Bouhmadi-L\'opez \\ \and 
Claus Kiefer \and 
Prado Mart\'in-Moruno
} 

\authorrunning{M.  Bouhmadi-L\'opez et al.} % if too long for running head

\institute{M. Bouhmadi-L\'opez \at
         	Department of Theoretical Physics, University of the
                Basque Country UPV/EHU, \\P.O. Box 644, 48080 Bilbao,
                Spain.\\	 
	IKERBASQUE, Basque Foundation for Science, 48011 Bilbao,
        Spain.\\                      
              \email{mariam.bouhmadi@ehu.eus}              
           \and
           C. Kiefer \at
              Institute for Theoretical Physics, University of
              Cologne, Z\"ulpicher Strasse 77, 50937 K\"oln,
              Germany.\\  
              \email{kiefer@thp.uni-koeln.de}
              \and
           P. Mart{\'i}n-Moruno \at
              Departamento de F\'isica Te\'orica and IPARCOS, Universidad
              Complutense de Madrid, 28040 Madrid, Spain.\\  
              \email{pradomm@ucm.es}   
}

\date{Received: date / Accepted: date}
% The correct dates will be entered by the editor

\maketitle

\begin{abstract}
Cosmological observations allow the possibility that dark energy is
caused by phantom fields. These fields typically lead to the occurrence
of singularities in the late Universe. We review here the status of
phantom singularities and their possible avoidance in a quantum theory
of gravity. We first introduce phantom energy and discuss its
behavior in cosmology. We then list the various types of
singularities that can occur from its presence. We also discuss the
possibility that phantom behavior is mimicked by an alternative theory
of gravity. We finally address the quantum cosmology of these models
and discuss in which sense the phantom singularities can be avoided.
% \PACS{PACS code1 \and PACS code2 \and more}
% \subclass{MSC code1 \and MSC code2 \and more}
\end{abstract}

\clearpage
%%%%%%%%%%%%%%%%%%%%%%%%%%%%%%%%
\section{Introduction}\label{sec:1}

The origin and fate of our Universe are central issues in
cosmology. Observations indicate that the expansion of the Universe is currently
accelerating and that it was probably also accelerating at an early
stage, the inflationary phase. 
The causes of those periods of accelerated expansion are unknown. It is also not known
whether the early-time and the late-time accelerations
have the same origin or not. It could be that the Universe was in a
highly symmetric (de~Sitter like) phase in the past and will return to
this in the future, cf. \cite{AAS79}. 
The source of the present acceleration 
could be a cosmological constant, an additional dynamical field, or a
modification of general relativity. The case of a
dynamical field is usually referred to as {\em dark 
  energy}.\footnote{Throughout this review \emph{dark
    energy} denotes a fluid that can be described by a field that is
  minimally coupled to gravity and 
  non-interacting with the standard model fields.  Otherwise we
  consider that the 
  field constitutes a genuine modification of general relativity and we have an
  alternative theory of gravity. Note that interactions within the dark sector are not excluded by this definition.} Most models for dark energy employ one or more scalar
fields.
     During the cosmic stages of acceleration
the {\em strong energy condition} $\rho+3p>0$, where $\rho$ is the
energy density and $p$ is the pressure,\footnote{We
  assume here that these fields can be described by an
  energy--momentum tensor that has the form of a perfect fluid. For
  homogeneous and isotropic geometries in general relativity, this is
  required for consistency.}  is violated by the fields that
drive the acceleration.
    Depending on whether the {\em null energy condition}
$\rho+p\geq0$ is fulfilled or not,\footnote{We use units where the
  speed of light $c=1$. In 
  SI units, this condition reads $\rho+p/c^2\geq0$, where both terms
  have the dimension ${\rm kg}/{\rm m}^3$. In these units it is seen
  that the second term is usually tiny compared to the first
  one. Taking air at sea level as an example, one has $\rho=1.29\ {\rm
    kg}/{\rm m}^3$ and $p/c^2=1.13\times 10^{-12}\ {\rm kg}/{\rm
    m}^3$. This means that pressures must be immensely large (and
  negative) in order to compensate for densities.} 
one talks about {\em standard} or
{\em phantom} fields.   Note that for both cases one must have a
positive energy density, $\rho>0$, since in cosmological spatially
flat scenarios only this implies that the square of the Hubble rate is
positive (as it must). There exist, of course, negative energy
densities in Nature, such as the Casimir energy, but they are usually
tiny. Phantom fields with $\rho<0$ are occasionally used as toy
models; see, for example, section~II.B in \cite{Dabrowski:2006dd}.

The current observational situation is such that both standard fields
and phantoms seem possible
\cite{Ade:2015xua,Aghanim:2018eyx,DES17,Bouali:2019whr}. Even more, it has been shown
that phantom models based on the ``vacuum-metamorphosis
model''\footnote{This model arises from taking into account loop
  corrections in the presence of a massive scalar field.} help to
alleviate the tension of the Hubble constant $H_0$ as inferred from Planck CMB data when interpreted within a $\Lambda$CDM
cosmological model and
local observational data \cite{DiValentino:2017rcr}. 
It is also possible that the current acceleration of the cosmic expansion
is a signal of the breakdown of general relativity at
cosmological scales and, therefore,
dark energy is just 
an effective description encapsulating modifications to the general
relativistic cosmic predictions. This possibility will be confronted
with upcoming observational data in the near future
\cite{DES17,Amendola:2012ys}.  We remind in this
regard that alternative theories of gravity are not only suitable  
to describe the late-time universe but also provide the model that
best fit the observations of the early Universe
\cite{Ade:2015xua,Aghanim:2018eyx,DES17,Starobinsky:1980te}.

This motivates the topic of our short review. Phantom fields are very
exotic in many respects, their energy densities increase with the
universe size, but they are a viable possibility from the
observational point of view. Among their exotic aspects is the
prediction of new types of singularities, notably the {\em big rip},
in which the observable Universe reaches infinite size in finite time. This
raises the question about the nature of these singularities, their
differences from the big bang singularity, and their fate in a quantum
theory of gravity. Such a theory is not yet available in final form,
but the question can be sensibly addressed in existing approaches
\cite{OUP}. Alternative theories of gravity can
describe an accelerating universe without the need of dark energy and,
therefore, they may also lead to phantom energy singularities.  
Thus, the formulation of a quantum alternative cosmology is a raising
field of study
\cite{Bouhmadi-Lopez:2016dcf,Albarran:2017swy,Alonso-Serrano:2018zpi,Bouhmadi-Lopez:2018tel,Albarran:2018mpg}. 

In standard cosmology, the classical singularity theorems by Penrose,
Hawking, and others, occupy a central place \cite{HE,HP}. 
How is a singularity defined? In the words of Hawking and Penrose
(\cite{HP}, p.~15): 
\begin{quote}
A spacetime is singular if it is timelike or null geodesically
incomplete but cannot be embedded in a larger spacetime.
\end{quote} 
Typical assumptions in the proof of those theorems are an energy
condition (such as the strong energy condition), a causality condition
(such as the absence of closed timelike curves), and a boundary or
initial condition (as the assumption that gravity is strong enough to
create a trapped 
region) \cite{Senovilla:2018aav}. 
A singularity defined in the above manner may or may not involve a
singularity in the curvature or matter properties (energy density,
pressure). One can also have curvature singularities without geodesic
incompleteness. Ellis {\em et al.} thus use the following more careful
definition (\cite{EMMbook}, p.~145):
\begin{quote}
We shall define a singularity as a boundary of spacetime where either
the curvature diverges \ldots or geodesic incompleteness occurs. The
relation between these two kinds or aspects of singularities is still
not fully clear; but often they will occur together.
\end{quote}
We shall adopt this point of view also here.

Our review is organized as follows.
In section~2, we discuss energy conditions in general and their status
for phantom fields in particular, emphasizing how to mimic a phantom behavior in alternative theories of gravity. Section~3 reviews and classifies
cosmic singularities. Section~4 is devoted to the quantum fate of
phantom singularities; we mainly discuss this question in quantum
geometrodynamics (with the Wheeler--DeWitt equation as its main
equation \cite{OUP}), but also address loop quantum cosmology and the
quantum cosmology of alternative theories of gravity.  Our section~5 concludes with a brief summary and outlook.

%%%%%%%%%%%%%%%%%%%%%%%%%%%%%%%%
%%%%%%%%%%%%%%%%%%%%%%%%%%%%%%%%
\section{Phantom energy}\label{sec:2}
A satisfactory theory of gravity should be
based on a relation between matter and geometry that provides us with
a description of how matter moves in a given spacetime and how it
affects its curvature.  
Although this relation is fixed by the theory, no restriction on the
nature of the material content is typically imposed by it. Therefore,
even in general relativity (GR), all possible four-dimensional
geometries seem to be allowed solutions. Restrictions are obtained by
causality conditions (e.g. avoidance of closed timelike curves) and
energy conditions (ECs). 
Imposing ECs, one attempts to restrict the
material content that should be taken into account. In particular,
they are assumptions made for the form of the stress--energy tensor in
agreement with our daily experience \cite{Matt,MP}. Hence, they can
provide us with some hints about the characteristics of the spacetimes
that are physically meaningful solutions of the theory. 
Nevertheless, the interpretation of the observational data currently
available in the framework of GR allows the existence of a fluid that
violates all the classical ECs as the most abundant substance in our
Universe. As we will summarize, such unorthodox phantom fluid may lead to instabilities when quantization of gravity is not considered. 
Moreover, those observational data may point to the
need of modifying GR at cosmological scales, where its validity has not been
properly tested but it is extrapolated. In the framework of alternative theories of gravity,
an effective phantom behavior can be described without introducing phantom
scalar fields. But one should mention that effects of dark energy (and
dark matter) can also be mimicked by infrared effects of quantum
gravity as they occur, for example, in the approach of asymptotic
safety \cite{RW04}.

%%%%%%%%%%%%%%%%%%%%%%%%%%%%%%%%
\subsection{Energy conditions}\label{sec:2.1}
In order to understand the gravitational effects felt by a family of
observers living in a given geometry, the Raychaudhuri equation can be
taken into account \cite{HE,Wald}. This is a purely geometric
relation in Riemannian geometry,
which describes the convergence or divergence of a congruence of
timelike (or lightlike) curves. For geodesic motion, it reads 
\begin{equation}
\label{Ray}
\frac{{\rm d}\theta}{{\rm d}s}=\omega_{ab}\omega^{ab}-
\sigma_{ab}\sigma^{ab}-\frac13\theta^2-R_{ab}V^aV^b,
\end{equation}
where  $R_{ab}$ is the Ricci tensor, $\omega_{ab}$ is the vorticity, $\sigma_{ab}$ is the shear,
$\theta$ is the expansion, and $V^a$ is the timelike unit vector tangent to
the congruence. This equation plays a central role in the proof of the
singularity theorems \cite{HE,Wald}.
Restricting ourselves to congruences with vanishing vorticity
(which have a tangent vector that can be expressed as a gradient of a
scalar field \cite{Abreu:2010eb}),\footnote{Vanishing vorticity is
  equivalent to the congruence being locally hypersurface orthogonal.}
 the Raychaudhuri equation implies
that those observers will get closer if
$V_a\,R^a{}_bV^b\geq0$. Therefore, the attractive character of gravity
is guaranteed if the \emph{timelike convergence condition} (TCC) is
satisfied, which requires $V_a\,R^a{}_bV^b\geq0$ for any timelike
vector $V^a$. Analogously, the focusing of null geodesics is implied by the
\emph{null convergence condition} (NCC), that is,
$k_a\,R^a{}_b\,k^b\geq0$ for any null vector $k^a$. 

As we have discussed in the introduction, 
a spacetime can be called singular if a freely falling observer
reaches the end of 
his or her path (although not only then). Therefore, as the TCC and NCC can be used
to prove timelike and null geodesics incompleteness under certain
circumstances \cite{HE,EMMbook}, they are of special interest to conclude the
singular character of some geometries. But we emphasize already here that
in the case of phantom energy, singularities can appear {\em because of} the 
violation of energy conditions. That is, if there is phantom energy,
the classical singularity theorems do not apply; however, this does
not imply that the geometry is geodesically complete nor that the
curvature invariants are finite throughout the spacetime. 

The first EC that we will discuss is intimately related with the
TCC. Indeed, it can be understood as the requirement of gravity to be
attractive in GR. This is the \emph{strong energy condition} (SEC),
which is mathematically formulated as
$V_a\left(T^a{}_b-\frac{1}{2}T\delta^a{}_b\right)V^b\geq0$,  where
$T^a{}_b$ are the $(1,1)$-components of the stress--energy tensor of
the matter fields and $T\equiv T^a{}_a$.   
This condition is equivalent to substituting the Einstein equations into the
TCC, so it makes sense only in the context of GR.
 For a perfect fluid, the SEC implies that $\rho+p\geq0$ and
$\rho+3p\geq0$, where $\rho$ and $p$ are, respectively, the energy
density and pressure of the perfect fluid as measured in its rest
frame. The Hawking and Hawking--Penrose singularity theorems assume the
TCC and thus the SEC in the framework of GR \cite{HE}. Hence, if the
TCC is satisfied during the cosmological evolution, we should accept a
singular origin of our Universe \cite{Wald}. Nevertheless, the SEC
(and the TCC) has to be violated  in cosmology when the expansion of
the Universe accelerates, both during the early inflationary phase and
right now. Therefore, it has been advocated that the SEC should be
abandoned \cite{Barcelo:2002bv}.  

The \emph{weak energy condition} (WEC) has a clear physical meaning
independent of the particular theory of gravity assumed. It requires
that the energy density measured by any observer must be
non-negative, that is, $V_a\,T^a{}_bV^b\geq0$ for any timelike vector
$V^a$. A perfect fluid satisfies the WEC if $\rho\geq0$ and
$\rho+p\geq0$. In the framework of GR, the WEC implies the
NCC. However, the NCC can also be satisfied by imposing only that the
WEC is fulfilled in the limit of null observers. This is the
\emph{null energy condition} (NEC), which requires that
$k_a\,T^a{}_b\,k^b\geq0$ for any null vector $k^a$; that is,
$\rho+p\geq0$ for perfect fluids.  

Regarding black hole geometries, the Penrose theorem and the Second Law of
black hole thermodynamics can be proven using the NCC \cite{Wald}. 
Moreover, as the TCC was necessarily violated during the early
inflationary phase of our Universe, a theorem pointing out the
existence of an initial singularity for open universes was
demonstrated under certain circumstances assuming only that the NCC is
satisfied \cite{Borde:1996pt}.\footnote{For closed universes, a similar
  theorem can be obtained assuming very strong requirements \cite{Borde:1996pt}. 
 These requirements can be violated (avoiding the conclusion of the theorem) for models of interest \cite{MolinaParis:1998tx} as,
  for example, in the
  emergent universe scenario discussed in \cite{EM04}, which describes an
  inflationary period without initial singularity.}
Therefore, the theorem requires just the fulfillment of the NEC in GR,
whereas the SEC could be violated \cite{Borde:1996pt}; see
\cite{Cattoen:2005dx} for a complete analysis of the ECs close to some
cosmological events of interest in
Friedmann--Lema\^{\i}tre--Robertson--Walker (FLRW) backgrounds. 

The \emph{dominant energy condition} (DEC) states that the energy
density measured by any observer is non-negative and propagates locally in a
causal way.
Therefore, by its very definition, the DEC implies the
WEC. Mathematically, the DEC requires that $V_a\,T^a{}_bV^b\geq0$ and
$F^aF_a\leq0$, with $F^a=-T^a{}_bV^b$ being the flux four-vector and $V^a$ any
timelike vector. This leads to $\rho\geq0$ and $|p|\leq\rho$ for a
perfect fluid. The zeroth law of black hole thermodynamics 
(i.e. that the surface gravity is constant over the horizon of a
stationary black hole) requires
the fulfillment of the DEC (\cite{Wald}, p.~334). 

According to the definitions of the ECs, it can be noted that, on one
hand, the DEC implies the WEC which leads to the NEC and, on the other
hand, fulfillment of the SEC implies that the NEC is satisfied (but
not necessarily the WEC); see, for example, Fig.~1 in
\cite{Carroll:2003st}. Therefore, violations of the NEC would lead
to violations of all the other mentioned ECs. 
In the classical realm, it is enough to consider a non-minimally 
coupled scalar field\footnote{This kind of couplings can be
  interpreted as a modification of GR (a scalar-tensor theory of
  gravity). 
It should be noted, however, that assuming a conformal coupling is
natural in some branches of physics.} to obtain such violations
\cite{Barcelo:2002bv}. Moreover, violations of all the ECs generically
appear when considering quantum vacuum states in semiclassical physics
(see the references in \cite{Matt} and \cite{MP}). 

In view of those violations, \emph{averaged energy conditions}, which
consist in integrating the ECs along timelike or null geodesics, were
taken into account. Although they only require the ECs to be satisfied
``on average'', there are also known violations of these conditions
\cite{Matt,MP}. A different approach is based on calculating \emph{quantum
  inequalities} \cite{Fewster:2012yh},   
which are bounds on the negativity of an average of the energy density.
Moreover, noting that negative energies in one region of the spacetime seem to 
be overcompensated by positive energies in other regions, 
the related \emph{quantum interest conjecture} was
formulated \cite{Ford:1999qv}   
(see also \cite{Fewster:2012yh,Abreu:2008dh} and references therein for developments).
Apart from that, following a local  point-wise
approach, nonlinear ECs were formulated
\cite{Martin-Moruno:2013wfa}. The most interesting example is the
\emph{flux energy condition} (FEC) that requires energy of any sign
to propagate in a causal way as seen by any observer. This condition
is satisfied in some semiclassical situations
\cite{MP,Martin-Moruno:2013wfa}. Furthermore, the semiclassical or
quantum ECs have been recently formulated, based on noting that
violations of the ECs are usually small. For example, the
\emph{quantum weak energy condition} (QWEC) demands that the energy
density measured by any observer should not be excessively negative
(so it can include, for example, the Casimir energy),
introducing a specific bound to  
quantify this claim  \cite{MP,Martin-Moruno:2013wfa}; see 
\cite{MP} and references therein for more details about these
extensions. 
 Whereas phantom energy violates the FEC, the QWEC and the QNEC have
 been used to minimize the violation of the ECs by phantom fields in
 cosmological \cite{Bouhmadi-Lopez:2014cca} or spherically symmetric
 solutions \cite{Bouhmadi-Lopez:2014gza}.

%%%%%%%%%%%%%%%%%%%%%%%%%%%%%%%%
\subsection{The phantom fluid}\label{sec:2.2}
\begin{figure}[h]
\centering
 \includegraphics[width=10cm]{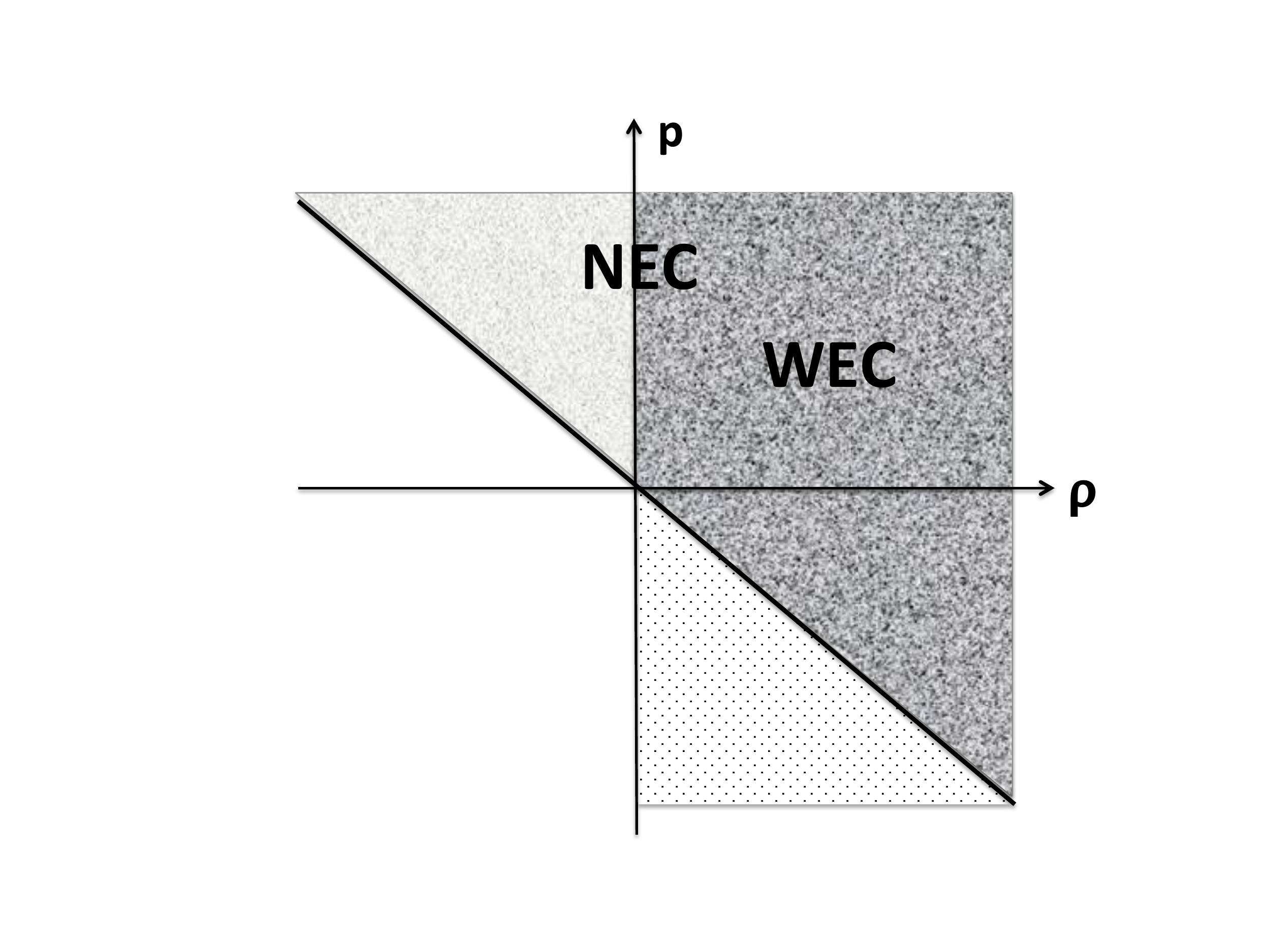}%\vspace{-0.5cm}%{\hspace{2cm}\includegraphics[width=5cm]{future.pdf}} 
\caption{This diagram shows the $p$-$\rho$-plane. 
    The whole shaded region corresponds to values of $\rho$ and $p$
    satisfying the NEC, whereas the WEC is satisfied only in the
    darker shaded region.  The dotted region corresponds to the
    phantom regime.} 
\label{figEC}
\end{figure}

Soon after the discovery of the accelerated expansion that our
Universe is currently undergoing, it was noted that dark energy could
violate not only the SEC, but also the NEC
$\rho+p\geq 0$. That is, the effective
equation of state parameter of dark energy $w:=p/\rho$, which needs to
be smaller than $-1/3$ to describe accelerated expansion for a general
relativistic universe, could be $w<-1$, cf. figure.\ref{figEC},
In this case, dark energy is called \emph{phantom energy} \cite{Caldwell:1999ew}.
Violations of the NEC would then not only be related with
semiclassical effects or appear in small quantities, but would be
relevant for classical fields leading to macroscopic effects.  
Nowadays, it is feasible that phantom energy is the most
abundant cosmological ingredient; indeed, we have $w=-1.03\pm0.03$ according
to the combination of Planck data and other astrophysical data
\cite{Aghanim:2018eyx} and $w=-1.00^{+0.04}_{-0.05}$ according to 
the combination of Planck 2015 data with DES performed by
the DES Collaboration \cite{DES17}.  In addition, as we mentioned in
the introduction, 
phantom models based on the vacuum-metamorphosis model helps to
alleviate the tension of $H_0$ as inferred from Planck CMB data when  interpreted within a $\Lambda$CDM
cosmological 
model and
local Hubble constant data \cite{DiValentino:2017rcr}.
The definition of phantom fields, stated above, does not impose a restriction on the
sign of $\rho$. It appears reasonable, however, to demand
$\rho>0$. Anyway, as can be seen from one of the
Friedmann--Lema\^{\i}tre equations,
\begin{equation}
  \dot{a}^2=-{\mathcal K}+\frac{8\pi G}{3}a^2\rho,
\end{equation}
where ${\mathcal K}$ is the curvature parameter, a negative
energy density is possible only for negatively curved Friedmann universes,
in which case (for $w<-1$) the universe expands from a
regular null hypersurface and contracts to another regular null
hypersurface \cite{HCI18}.\footnote{Ref.~\cite{HCI18} contains a
  complete classification of all FLRW solutions with equation of state $p=w\rho$.}

%%%%%%%%%%%%%%%%%%%%%%%%%%
%%%%%%%%%%%%%%%%%%%%%%%%%%

The potential existence of such exotic fluids challenges our
understanding of nature.  
The energy density of phantoms increases with the cosmic
expansion and its dominance leads to a superaccelerating universe (a
universe with a cosmic acceleration larger than that of a de Sitter
model), whose scale factor may even blow up at a finite time (see
section \ref{sec:3.1.1} for more details). Moreover, a fluid violating
the NEC could allow the existence of exotic geometrical objects such as
wormholes \cite{Matt}; therefore, if dark energy is of phantom nature,
those objects may exist in our Universe with macroscopic size 
\cite{Sushkov:2005kj,Lobo:2005us}. 

The most important potential shortcoming of phantom
energy concerns its stability. 
The simplest fundamental description of a phantom fluid 
is in terms of a scalar field that is minimally coupled with gravity 
and has a canonical kinetic term with the ``wrong" sign
\cite{Caldwell:1999ew}.
 Even if such a  ``ghost field'' does not interact classically with other
 fields, it will interact at least with the graviton.\footnote{See,
   for example, \cite{sbisa} for a review on classical and quantum
   ghost fields.}
 This will lead to a quantum instability of the vacuum 
if we do not consider the theory as an effective description valid 
only under a given cutoff
\cite{Carroll:2003st,Cline:2003gs}.\footnote{It should be emphasized 
  that the potential ghost instability appears when the quantization of
  gravity is not considered together with that of the field (more
  comments along these lines will be included in section
  \ref{sec:4}).}   
The ghost instability may be avoided when considering a non-canonical
kinetic term for the scalar field. From
studying perturbations of such ``k-essence fields'' around a FLRW 
background solution, one can conclude, however, that if the field violates the
NEC without introducing a ghost, the speed of sound for the
perturbations is imaginary \cite{Creminelli:2008wc}, that is, the
system has a gradient instability. 
More complicated Lagrangians for a scalar field, which are still
minimally coupled to gravity but include second derivatives of the
field in such a way that the field equations remain second order,
may lead to violations of the NEC without instabilities or a
superluminal speed of sound  at least in FLRW backgrounds 
\cite{Rubakov:2014jja,Rubakov2}.  Therefore, even if the potential
semiclassical instability of phantom energy could lead us to abandon
the consideration of this exotic fluid, we prefer to keep an open mind
until its fundamental (classical and quantum) field description is
unveiled, at least as long as the observational data allow this
possibility.

%%%%%%%%%%%%%%%%%%%%%%%%%%%%%%%%
\subsection{Effective phantom behavior}\label{sec:2.3}
The potential existence of a dark fluid with anti-gravitational
properties (that is, violating the SEC) as the most abundant
cosmological ingredient in our Universe is a result of interpreting
the observational data in the framework of GR under the assumption of 
the cosmological principle.\footnote{The case of a cosmological
  constant,  understood as a fluid with $p=-\rho$,  also violates the SEC, but saturates the NEC (so it can be considered as a limiting case between phantom and dark energy).}  
However, it is possible that the current
accelerated expansion of our Universe is just a signal of the need of
modifying GR at large scales. Alternative theories of gravity
were already suggested as well-motivated candidates to describe
gravitational phenomena at high energies or short scales
\cite{CFbook}. At present, the Starobinsky
model of cosmic inflation is considered to be a plausible
description of the early Universe  (\cite{Ade:2015xua}, Sec.~6).
Therefore, on the one hand, one could
think that the potential existence of a phantom fluid (whose energy density grows with time as a results of the NEC violation) will drive the
Universe again to a high-energy regime where modifications of GR might
be necessary. On the other hand, one can go a step further and
interpret that GR may break down in both the ultraviolet cosmic regime
(during the early cosmological inflationary phase) and the infrared cosmic
regime (recently in the cosmological evolution), with GR being only valid
in a limited intermediate regime of the cosmic evolution. 
This interpretation suggests the need for considering alternative
theories as the fundamental of gravitational phenomena. Thus, following this second point of view, the apparent
existence of a dark fluid violating the SEC may indicate that we
are entering a regime for which the appropriate gravitational theory
deviates from GR significantly. This interpretation would appear
more justified if the hypothetical fluid violated the WEC, explaining the
counterintuitive nature of such a fluid in the classical realm. 

In order to understand in a quantitative way this idea, let us note
that alternative theories of gravity typically introduce higher-order
curvature invariants and/or new (scalar, vector, or tensor)
gravitational fields non-minimally coupled to gravity in the
Lagrangian. Therefore, the modified Einstein equations of a large
class of alternative theories of gravity can be written as
\cite{Capozziello:2014bqa} 
\begin{equation}\label{mE}
g(\Psi^i)\left(G_{ab}+H_{ab}\right)=8\pi G\, T^{\rm (m)}_{ab},
\end{equation}
where $G_{ab}$ is the Einstein tensor, $T^{\rm (m)}_{ab}$ is the
stress--energy tensor associated with the material content (we assume
that we have here no dark energy component), $H_{ab}$ encapsulates
additional gravitational terms of geometrical nature or involving the
new fields, and $g(\Psi^i)$ takes into account potential modifications
of the gravitational coupling with $\Psi^j$ denoting curvature
invariants and gravitational fields. It is easy to rewrite
(\ref{mE}) as 
\begin{equation}\label{mE2}
G_{ab}=8\pi G_{\rm eff} \left[T^{\rm (m)}_{ab}+T^{\rm (eff)}_{ab}\right],
\end{equation}
with
\begin{equation}\label{eff}
G_{\rm eff}= \frac{G}{g(\Psi^i)}\qquad{\rm and}\qquad T^{\rm
  (eff)}_{ab}=-\frac{g(\Psi^i)}{8\pi\,G}H_{ab}.
\end{equation}
Thus,
\begin{equation}\label{tot}
T^{\rm (tot)}_{ab}=T^{\rm (m)}_{ab}+T^{\rm (eff)}_{ab}.
\end{equation}
The SEC, which is the requirement of the TCC being satisfied in GR, makes no
sense once we consider a different theory of gravity. One can
formulate a generalization of this condition requiring the TCC to be
satisfied in the new theory \cite{Capozziello:2014bqa}, obtaining an
inequality involving both stress--energy tensors contracted with a
timelike vector and their traces. Following this procedure, one can
study violations of the TCC induced by a non-vanishing effective
stress-energy tensor.  

Whereas $T^{\rm (m)}_{ab}$ is associated with real matter
fields, $T^{\rm (eff)}_{ab}$ is of purely gravitational origin
comprising geometric terms and/or extra gravitational
fields. Therefore, one can impose that the WEC be satisfied for
the matter stress--energy tensor, but there is no reason why the energy
density associated with the effective stress--energy tensor should be
positive, since this tensor just encapsulates the modification with
respect to GR \cite{MP,Barcelo:2002bv}. Indeed, in some theories one
can easily obtain $T^{\rm (eff)}_{ab}k^ak^b<0$; see, for example,
\cite{Baccetti:2012re,Albareti:2012va}. Following this
spirit, in alternative theories of gravity one can obtain geometries
where the NCC is violated supported by a material content with a
positive energy density as measured by any observer. 
 That is, one can violate the NCC without violating the energy
 conditions in alternative theories of gravity.\footnote{Notice as
   well that it is possible to have a phantom-like behavior, that is,
   to define an effective energy density, $\rho_{\textrm{eff}}$, such
   that $\dot\rho_{\textrm{eff}}>0$ in an expanding universe if the
   spatial geometry is spherical and
   $\rho_{\textrm{eff}}:=\Lambda-\frac{1}{a^2}$ where $\Lambda$ is a
   cosmological constant \cite{Clarkson:2007bc}.} 
 This is the 
ultimate reason for the possibility of describing phantom cosmologies
in the framework of alternative theories of gravity without the
introduction of exotic fluids; see, for example,
\cite{Nojiri:2010wj,Ludwick17}.

%%%%%%%%%%%%%%%%%%%%%%%%%%%%%%%%%%%%%%%%%%%%%%%%%%%%%%%%%%%%%%%%%%
\section{Cosmic singularities}\label{sec:3}

The characterization of singularities is a fascinating and at
the same time a difficult task to tackle in a relativistic
gravitational theory, no matter if it is GR or any
extension of it. To give a metaphor: ``How
    abrupt can the tip of a mountain appear? The answer depends on the
    shape of its 
    summit or on the different ways of reaching and/or unreaching
    the summit.'' Singularities are a familiar concept in various
    branches of physics, especially in fluid dynamics; see, for
    example, \cite{Moffatt} and the references therein.
Within a relativistic gravitational theory, the
shape of the mountain can be characterized by its geometry and
therefore by its curvature, while the ways of reaching its summit can be
described through parametric curves, the timelike and
lightlike geodesics being the easiest way to characterize the paths, which
describe freely falling object that do not experience any
acceleration.  

As we have already stated in the introduction, a singularity can be
defined as a boundary of spacetime where 
either the curvature is ill-defined (e.g. it or some of its
derivatives diverge) or geodesic incompleteness occurs \cite{EMMbook}. 
In the light of these criteria, we will next summarize work carried out in
the context of dark energy singularities and abrupt events.

\subsection{Classification of singularities}\label{sec:3.1}

First of all, we would like to recall that within GR
there is a direct and bijective (linear) relation between the energy
density and pressure of the effective (total) fluid filling a FLRW
universe and the (squared) Hubble rate and its cosmic time
derivative:\footnote{We note that this relation follows from the
  Raychaudhuri equation (\ref{Ray}) in the Friedmann limit if
  Einstein's equation hold.}
\begin{equation}
H^2+\dot{H}=-\frac{4\pi G}{3}(\rho+3p).
  \end{equation}
This relation does no longer need to exist in alternative theories of
gravity. For this reason, we will characterize
cosmological singularities by the Hubble rate and its cosmic time
derivative(s), which define univocally the scalar curvature of spacetime,
rather than through the total energy density and pressure of  matter
filling the Universe. 
We will as well distinguish between {\em cosmic singularities} that happen
at a finite cosmic time and {\em abrupt events} that happen at an infinite
cosmic time. 

\subsubsection{Metric classification}\label{sec:3.1.1}

\begin{enumerate}
\item {\em Cosmic curvature singularities (at a finite time): GR and beyond}

\vskip 2mm

The following classification is a generalization of the scheme
proposed in \cite{Nojiri:2005sx}.

\begin{description}

\item[{\bf Type 0-a}] A big bang singularity is a past singularity that
  takes place at vanishing scale factor where the Hubble rate and
  its cosmic time derivative diverge \cite{EMMbook}.  

\item[{\bf Type 0-b}] A big crunch singularity is a future singularity that
  takes place at vanishing scale factor where the Hubble rate and
  its cosmic time derivative diverge \cite{EMMbook}. For a FLRW
  universe, this singularity is usually present in a universe which is spatially
  closed and filled with matter satisfying the strong energy
  conditions (but see \cite{Barrow86} for a counterexample).  

\item[{\bf Type I}] A big rip singularity takes place at finite cosmic
  time with infinite scale factor, where the Hubble parameter and
  its cosmic time derivative diverge
  \cite{Starobinsky:1999yw,Caldwell:2003vq,Caldwell:1999ew,Carroll:2003st,Chimento:2003qy,Dabrowski:2003jm,GonzalezDiaz:2003rf,GonzalezDiaz:2004vq,Dabrowski:2006dd,Albarran:2015tga}. The
  occurrence of this singularity is intrinsic to phantom dark energy.   

\item[{\bf Type II}] A sudden singularity takes place at finite cosmic time
  with finite scale factor, where the Hubble 
parameter remains finite but its cosmic time derivative diverges
\cite{Barrow86,Barrow:2004xh,Nojiri:2005sx,Gorini:2003wa}. It is also
known as a big brake \cite{Gorini:2003wa,Kamenshchik:2007zj,Nick2016} if it
takes place in the future with an infinite deceleration or a big
d\'emarrage  \cite{BouhmadiLopez:2009pu}, named also big boost
\cite{Barvinsky:2008rd}, if it happens in the past with an
infinite acceleration. The latter case is realized by a generalized
Chaplygin gas \cite{BouhmadiLopez:2009pu,BouhmadiLopez:2007qb}, see
also section~4.2.2 below. 

\item[{\bf Type III}] A big freeze singularity takes place at finite
  cosmic time with finite scale factor, where the Hubble parameter
  and its cosmic time derivative diverge
  \cite{Nojiri:2005sx,BouhmadiLopez:2009pu,BouhmadiLopez:2007qb,Nojiri:2004pf,BouhmadiLopez:2006fu,Nojiri:2005sr}. This
  singularity has also been named finite scale factor singularity in
  \cite{Dabrowski:2009pc}. A phantom generalized Chaplygin gas can
  induce a big freeze in the future. 

\item[{\bf Type IV}] This singularity takes place at finite cosmic time
  with finite scale factor, where the Hubble parameter and its
  cosmic time derivative remain finite, but higher cosmic time
  derivatives of the Hubble parameter diverge
  \cite{Nojiri:2005sx,Nojiri:2004pf,BouhmadiLopez:2006fu,Nojiri:2005sr,Nojiri:2008fk,Bamba:2008ut,Bouhmadi-Lopez:2013tua}. They 
  have been also called  generalized sudden singularities
  \cite{Dabrowski:2013sea}. We emphasize that the
  scalar curvature is well defined not only through its finiteness but
  also through its differentiability. Therefore, a type IV singularity
  can be seen in GR as an event where the curvature is
  not (or not completely) differentiable, that is, it is not a $C^{\infty}$
  function. Beyond GR this singularity can affect the equations of motion through derivatives of the scalar curvature, as, for example, in
metric $f(R)$ models \cite{Nojiri:2008fk}.

\item[{\bf Type V}] A $w$-singularity takes place at finite cosmic time
  with finite scale factor, where the Hubble parameter vanishes and
  its cosmic time derivative is finite in such a way that the
  barotropic index of the fluid dominating the cosmic dynamic blows up
  \cite{Dabrowski:2009kg}. Notice that in a type IV singularity this
  behavior does not necessarily happen, but it is not
  forbidden. While within GR this behavior will
  not affect the boundedness of the curvature, it can affect its
  differentiability as for type~IV. 
In addition, at first
  order in the cosmological perturbations, the unboundedness of $w$ can
  have an effect on the evolution equations of the gravitational
  potential when approaching the singular point (cf. Eq. (3.16) in
  \cite{Albarran:2016mdu}).   

\item[{\bf Type VI}] A $Q$-singularity takes place at finite cosmic time
  in models with interacting dark energy and dark matter in which the
  interacting term $Q$ blows up
  \cite{BeltranJimenez:2016dfc,Chimento:2015gga}. While this does not
  mean any geometrical issue at the background level, it might
  highlight possible instabilities at the perturbative level. More precisely, what happens is that a divergence in $Q$ may lead to a blow up 
  of the time derivative of the equation of state and this on its own can result in a very large speed of sound of dark energy and consequently in an instability of the cosmological perturbations. 
 Therefore,  as in a
  $w$-singularity, a $Q$-singularity can be seen at the perturbative
  level. 

\end{description}

\item {\em Abrupt cosmic events\footnote{We will be referring to
      abrupt events as those cosmic curvature singularities that take
      place at an infinite cosmic time and where all  
the structure in the Universe can be destroyed in the far future (at a
finite cosmic time from now.). Therefore, this definition excludes the
pseudo-rip given that it corresponds to a mild event. In fact, it
corresponds to a model where the Hubble parameter 
  increases monotonically and
reaches a constant value at infinite cosmic time and scale factor
\cite{Frampton:2011aa}.}} 

\vskip 2mm

\begin{description}

\item[{\bf Little rip}] This takes place at an infinite cosmic time with
  infinite scale factor, where the Hubble rate and its cosmic time
  derivative diverge. 
It can visualized as a big rip sent towards an infinite cosmic time 
\cite{Ruzmaikina1970,Barrow:1990vx,Nojiri:2005sx,Stefancic:2004kb,BouhmadiLopez:2005gk,Frampton:2011sp,Brevik:2011mm,Bouhmadi-Lopez:2013nma,Albarran:2016ewi}. 
It is worth mentioning that the first time this behavior was found
was in an alternative theory of gravity 
\cite{Ruzmaikina1970}. Mathematically equivalent solutions can be
obtained in some specific inflationary models where the expansion is
driven by a perfect 
fluid with bulk viscosity \cite{Barrow:1986yf,Barrow:1988yc}. 

\item[{\bf Little sibling of the big rip}] This takes place at infinite
  cosmic time with infinite scale factor, where the Hubble rate
  diverges, but its cosmic time derivative remains finite. Such a
  behavior can happen precisely because it takes place at an
  infinite cosmic time \cite{Bouhmadi-Lopez:2014cca,Albarran:2015cda}.  

\end{description}

We emphasize that these abrupt events are intrinsic to phantom dark energy.

\end{enumerate}

\subsubsection{Affine classification}

An alternative way of characterizing a spacetime singularity is by
analyzing its causal curves, in particular its geodesics. We can
therefore say that a spacetime has a singularity if there are causal
curves that are incomplete
\cite{EMMbook,FernandezJambrina:2004yy,FernandezJambrina:2006hj}. We
note that while the definition of cosmological singularities used
in the previous subsection is independent of observers, the
current one will depend on them. A timelike
geodesic, for example, might be incomplete, while a lightlike geodesic might be
complete. This is what happens in a big rip model with a barotropic
equation of state where $-5/3<w<-1$; that is, under these conditions a
lightlike geodesic will never reach the big rip, as it would take it
an infinite affine proper time, while a timelike geodesic will
hit the big rip in a finite affine proper time
\cite{FernandezJambrina:2006hj,HCI18}. 

We start by recalling the (crucial) causal geodesic equation in a
spatially flat FLRW universe, which at first order of a power
expansion reads \cite{FernandezJambrina:2006hj}: 
\begin{equation}
\dot{t}=\sqrt{\delta+\frac{P^2}{a^2(t)}}, \quad P={\textrm{constant}},
 \quad \delta=0,\pm1. 
\end{equation}
As pointed out in
\cite{FernandezJambrina:2004yy,FernandezJambrina:2006hj} (see also
\cite{FernandezJambrina:2007sx}),  
this equation is well defined for any  finite non-vanishing value of
the scale factor. A careful analysis of causal geodesics in dark
energy dominated universes was carried out in
\cite{FernandezJambrina:2004yy,FernandezJambrina:2006hj}  
under the assumption that the scale factor $a(t)$ allows a generalized
Puiseux expansion, that is, a generalized Taylor expansion where the
exponents are real and not necessarily  natural numbers and where
there is always a minimum finite exponent from which the expansion is
made around the singular event. This analysis was extended to some
inspired modified theories of gravity \cite{FernandezJambrina:2008dt},
$w$-singularities \cite{FernandezJambrina:2010ck}, and more recently
to  $Q$-singularities 
\cite{BeltranJimenez:2016dfc}. In all these works, it was concluded
that cosmic curvature dark energy singularities would be reached by
causal geodesics in a finite cosmic time with the exception of
lightlike geodesics in case of a barotropic equation of state with $ -5/3
< w < -1$.  

The next question to address is: how strong are these singularities? In
the previous subsection, the strength of the singularity was characterized
by the divergence of the Hubble rate and its cosmic
time derivatives at the singularity, that is, the degree of divergence for
quantities that characterize or define the curvature as well as its
analyticity.  As we are dealing with geodesics, the
simplest way is to analyze scalars constructed from the contraction
between the Ricci tensor and the velocity of the observer. This is
precisely what the criteria by Tipler and Kr\'olak achieve, see below.

Before proceeding further, we recall that the idea of a strong
singularity was first introduced by Ellis and Schmidt at the end of
the seventies \cite{Ellis:1977pj}. So far, we have assumed that the
observers are point-like objects but in fact they can have a finite
volume and therefore they are subject to tidal forces. With this
in mind, it is understood that a singularity is strong if tidal
forces imply a wrecking of the object that is heading towards the
singularity. Tipler and Kr\'olak independently proposed criteria to quantify
the effect of tidal forces on such an object. Let us have a closer
look at them.

{\em Tipler criterion}: Following Tipler \cite{Tipler:1977zza}, a
singularity is strong if the volume characterizing the
object\footnote{We will not enter into the technicality of how the
  volume is defined; instead we refer to
  \cite{Tipler:1977zza,FernandezJambrina:2004yy,FernandezJambrina:2006hj}
  for further details.} vanishes when the geodesics reach the
singularity.  
This criterion was adapted to expanding FLRW universes filled with
phantom dark energy
\cite{FernandezJambrina:2004yy,FernandezJambrina:2006hj} and implies
there that this volume diverges.
In fact, for an expanding FLRW universes filled with phantom dark
energy, the spacetime contains a strong singularity at $\tau_0$ if
\cite{ClarkeKrolak,FernandezJambrina:2006hj} 
\begin{equation}
\int_0^{\tau}{\rm d}\tau'\int_0^{\tau'}{\rm d}{\tau''}R_{ij}u^iu^j,
\end{equation} 
diverges when $\tau$ approaches $\tau_0$. Here,
$u^i$ stands for the geodesic four-velocity with affine parameter $\tau$. This
definition applies to lightlike and timelike geo\-de\-sics. 

{\em Kr\'olak criterion}: Following Kr\'olak \cite{Krolak1986}, a
singularity is strong if 
the proper-time derivative of the volume of the object heading towards the
singularity is negative. Therefore,
this criterion is less restrictive than the Tipler one. The
Kr\'olak criterion was also applied to expanding FLRW universes
filled with phantom dark energy
\cite{FernandezJambrina:2004yy,FernandezJambrina:2006hj} and implies
there that the cosmic derivative of the volume must be positive.  
In fact, for an expanding FLRW universe filled with phantom dark
energy, the spacetime contains a strong singularity at $\tau_0$ if
\cite{ClarkeKrolak,FernandezJambrina:2006hj} 
\begin{equation}
\int_0^{\tau}{\rm d}\tau'R_{ij}u^iu^j
\end{equation}
diverges when $\tau$ approaches $\tau_0$. Again, $u^i$ stands for
the geodesic velocity with respect to an affine parameter $\tau$. This definition applies
to lightlike as well as timelike geodesics. 

Applying these criteria
\cite{FernandezJambrina:2004yy,FernandezJambrina:2006hj,FernandezJambrina:2008dt,FernandezJambrina:2010ck,BeltranJimenez:2016dfc}, 
it was shown that all dark energy cosmic singularities are weak
with the exceptions of: (i) the big rip, which is a strong singularity,
and (ii) the big freeze, for which the Kr\'olak  and Tipler criteria give
opposite results: while for the former the singularity is strong, for
the latter it is weak; therefore, no conclusions can be
drawn in this case. Given that it has been shown
that all bounded structure of the universe facing a big freeze in the
future will be destroyed \cite{Bouhmadi-Lopez:2014jfa}, we believe
that the Kr\'olak criterion is more appropriate. 

We finally like to emphasize two things.
First, by analyzing the geodesics it is possible to find spacetime
pathologies which are hidden in the definition of the spacetime
curvature. This is the case for \textit{directional singularities} 
where the Ricci curvature vanishes at the singularity, but where the
projection of the curvature along some causal geodesics diverges
\cite{FernandezJambrina:2007sx}. Second, one may also apply the
generalized geodesic deviation equation \cite{PO16} when investigating
the approach to singularities. 

\subsection{Phantom singularities and alternative theories of gravity}
Among the cosmic singularities and abrupt events discussed above,
only some of them require a violation of the
null convergence condition (at least from a phenomenological point of
view), that is, $p^{\textrm{(tot)}}+\rho^{\textrm{(tot)}}<0$
(cf.~equation (\ref{tot})); 
those correspond to
the big rip, pseudo-rip, little rip, and 
little sibling of the big rip. If, in addition, we impose that the
singularity takes place in the future, the big freeze will also
be included.

So far, we have assumed that a dark energy singularity
arises from a perfect fluid within GR. But this is not the only
possibility. Such a singularity could also arise from 
modified gravity, for example from an action containing a function of
the Ricci scalar, $f(R)$, within the
metric formalism. As is well known, a perfect fluid with a constant
equation of state (or more complicated variations) can be perfectly modeled
in this setup; see, for example, \cite{Morais:2015ooa} and the 
references therein. Indeed, if $f(R)$
is a linear combination of the following form \cite{Morais:2015ooa}: 
\begin{equation}
f(R)=C_+R^{\beta_+} + C_-R^{\beta_-}, \quad C_\pm=\rm{constant},
\end{equation}
where 
\begin{equation}
\beta_\pm=\frac12\left\{1+\frac{1+3w}{6(1+w)}
\pm\sqrt{\frac{2(1-3w)}{3(1+w)}+\left[1+\frac{1+3w}{6(1+w)}\right]^2}\right\}, 
\end{equation} 
we will find for a FLRW universe an expansion that is equivalent to
that of a FLRW universe in GR with a perfect fluid satisfying
$p=w\rho$ with constant $w$. In particular, one can mimic a perfect fluid $w<-1$ at the background level. Other $f(R)$ metric theories could
induce a sudden, big freeze, or type IV singularity, given that any kind
of modified generalized Chaplygin gas can (i) be mimicked by a metric
theory and (ii) is known to induce the singularities listed above (for
certain choices of the parameters of the model). In summary, what we
want to stress is that the phantom nature of a dark energy could be
rooted in some alternative theory of gravity (as already argued in
section \ref{sec:3.1} above), and we have given here a simple example within the
framework of $f(R)$ gravity.  

Before ending this subsection, we want to emphasize two things.
First, even in the absence
of true phantom matter in GR, we can sometimes define an
effective dark energy that grows as the Universe expands. Such
a description might pretend the existence of a phantom field, while
it could be simply an effect of non-flat spatial curvature
\cite{Clarkson:2007bc} or of extra dimensions
\cite{Chimento:2006ac,BouhmadiLopez:2008nf}. 
Second, phantom matter does not
necessarily imply the existence of a future singularity (`doomsday'). 
For example, an equation of state
that violates the NEC but that asymptotically
approaches the equation of state corresponding to a cosmological
constant might simply lead to a de~Sitter universe 
\cite{BouhmadiLopez:2004me}. 

After having classified the different types of phantom singularities,
we next address the issue of how to cure them or at least how to
smooth them. The obvious answer is to consider a quantum treatment, as
we will do in the next section. We expect as well that GR will be
modified when approaching those singularities, 
that is, we expect a semiclassical regime when approaching the occurrence
of the singularity. We will consider this second option in the rest of
this subsection. 

Given that phantom singularities happen at very high energies and at
very late time, ultraviolet (UV) and/or infrared (IR) corrections to
GR could occur and a simple approach to describing those
corrections is within the framework of alternative theories of
gravity. We end with two examples that characterize this idea.

First, five-dimensional braneworld models with an induced gravity term on
  the brane and a Gauss-Bonnet contribution to the bulk action include
  IR and UV corrections to GR in a natural way
  \cite{Brown:2005ug}. Within this setup it has been shown that the
  big rip singularity can be alleviated and substituted by a sudden
  singularity \cite{BouhmadiLopez:2009jk}. 

Second, the Eddington-inspired-Born Infeld (EiBI) model has attracted
  some attention lately \cite{Banados:2010ix}. It corresponds to
  a Palatini theory, that is, to a theory where the connection that defines
  the curvature is not the Christoffel symbols of the physical
  metric. It has the virtue of removing the big bang singularity
  \cite{Banados:2010ix} and smoothing some phantom dark energy
  singularities \cite{Bouhmadi-Lopez:2014jfa}.

%%%%%%%%%%%%%%%%%%%%%%%%%%%%%%%%%%%%%%%%%%%%%%%%%%%%%%%%%%%%%%%%

\section{Quantum fate of classical singularities}\label{sec:4}

\subsection{Criteria for singularity avoidance}\label{sec:4.1}

In GR, we have well defined criteria for singularities
(such as geodetic incompleteness), and we have rigorous singularity theorems
\cite{HE,HP,EMMbook}. None of these are available in quantum
gravity. The main reason for this is, of course, that no such theory
exists in final form \cite{OUP}. But even given particular approaches,
we are still far from presenting exact definitions of
singularities and their potential avoidance in quantum gravity. 
Nevertheless, heuristic criteria for singularity avoidance exist and
have been successfully applied to cosmological models.  

There are various approaches to quantum gravity, but the most
straightforward and conservative one is the attempt to directly quantize
GR. For this purpose, one can formulate GR in
canonical form to arrive at a picture of wave functions in
configuration space \cite{OUP}. In the following, we shall concentrate
on quantum geometrodynamics, where the configuration space is the
space of all three-geometries $^{(3)}{\mathcal G}$ called superspace,
but we also include a brief discussion 
of loop quantum cosmology. In quantum geometrodynamics, the central
equation is the Wheeler--DeWitt equation
\begin{equation}
\label{wdw}
H\Psi=0,
\end{equation}
where $\Psi$ is the wave functional on the configuration space of
three-geometries and non-gravitational fields.\footnote{In the full
  theory, the Wheeler--DeWitt equation is complemented by the
  diffeomorphism constraints, which imply that $\Psi$ is invariant
  with respect to infinitesimal three-dimensional coordinate transformations.}  

In classical GR, singularities can occur at particular values of the
three-geometry. In his pioneering paper \cite{DeWitt67}, DeWitt
focused attention on three-geometries with vanishing volume
(corresponding, in cosmology, to the big bang and big crunch
singularities). Calling a singular boundary of configuration space
${\mathcal B}_Q$, DeWitt adopts the following criterion
(\cite{DeWitt67}, p.~1129):
\begin{quote}
The fact that ${\mathcal B}_Q$ is not the empty set \ldots is not
necessarily embarrassing to the quantum physicist, for he may be able
to dispose of it by simply imposing, on the state functional, the
following condition:
\begin{equation}
\label{DeWitt}
\Psi\left[^{(3)}{\mathcal G}\right]=0 \quad {\rm for\  all}\
^{(3)}{\mathcal G}\  {\rm on}\  {\mathcal B}_Q.
\end{equation}
{\em Provided it does not turn out to be ultimately
  inconsistent},\footnote{For example, by allowing only the trivial
  solution $\Psi\equiv0$ (our comment).}
this condition \ldots yields two important results. Firstly, it makes
the probability amplitude for catastrophic 3-geometries vanish, and
hence gets the physicist out of his classical collapse
predicament. Secondly, it may permit the Cauchy problem for the ``wave
equation''\footnote{This equation is the Wheeler--DeWitt equation (\ref{wdw}).}
 \ldots to be handled in a manner very similar to that of
the ordinary Schr\"odinger equation.
\end{quote}
This ``DeWitt criterion'' is based on the heuristic idea that wave
functionals in quantum gravity can be related, in some sense, to
probability amplitudes as in ordinary quantum theory. This is not
obvious, since (\ref{wdw}) does not contain any external time
parameter, so the main argument for the probability interpretation
(conservation of probability in time, that is, unitarity) is not
applicable. 

The DeWitt criterion $\Psi\to 0$ can only serve as a sufficient, not as a
necessary condition. Let us recall, for example, that the solution of
the Dirac equation for the ground state of the hydrogen atom diverges
at the origin,
\begin{displaymath}
\psi_0(r)\propto (2mZ\alpha r)^{\sqrt{1-Z^2\alpha^2}-1}{\rm e}^{-mZ\alpha r}\quad
\stackrel{r\to 0}{\longrightarrow} \quad\infty.
\end{displaymath}
Nevertheless, this state is not singular, since the integral
$\int{\rm d}r\ r^2\vert\psi_0\vert^2$ remains finite. 

Another criterion adopted in quantum cosmology is the breakdown of the
classical approximation when approaching the singularity
\cite{Dabrowski:2006dd}. This means that wave packets necessarily
disperse in this region and that the classical singularity
theorems do not apply. The general situation of late-time
singularities in quantum cosmology is carefully reviewed in \cite{Sasha13}.

 The vanishing of the wave function as a criterion for singularity
 avoidance is also adopted in other contexts. The
 authors of \cite{KKN09} discuss the situation for eleven-dimensional
 supergravity. They employ a cosmological billiard description near a
 spacelike singularity and find that the wave function solution of the
 corresponding Wheeler--DeWitt equation approaches zero there. Such a
 behavior also occurs in models of gravitational collapse and can
 there be interpreted as the avoidance of a black-hole singularity
\cite{HK01,MG15}. In the latter case, $\Psi=0$ at the position of the
classical singularity follows from the unitary evolution with respect
to a dust proper time. 

%%%%%%%%%%%%%%%%%%%%%%%%%%%%%%%%
\subsection{Quantum phantom cosmology}\label{sec:4.2}

In most investigations, restriction is made to FLRW universes. In this
case, the Wheeler--DeWitt equation reads (with units $2G/3\pi=1$)
\cite{OUP,Calcagni}
\be
\lb{mini}
\frac{1}{2}\left(\frac{\hbar^2}{a^2}\frac{\partial}{\partial a}
\left(a\frac{\partial}{\partial a}\right)-\ell\frac{\hbar^2}{a^3}
\frac{\partial^2}{\partial\phi^2}-{\mathcal K}a+\frac{\Lambda a^3}{3}+
2a^3V(\phi)\right)\psi(a,\phi)=0 ,
\ee
where $a$ is the scale factor, ${\mathcal K}=0,\pm1$, $\Lambda$ the
cosmological constant, 
and $\phi$ a homogeneous scalar field mimicking matter; $\ell$ can
assume the values $+1$ (ordinary field) and $-1$ (phantom field).
The factor ordering has been chosen in order to achieve covariance in
configuration space (Laplace--Beltrami factor ordering). We note that
 the scale factor $a$ can also be interpreted as a kind of
 ``phantom''. In the full Wheeler--DeWitt equation, the kinetic term
 related with the (local) volume of the three-geometry is negative, a
 feature that can be related to the attractivity of gravity
 \cite{GK94}.

\subsubsection{Examples with big rip, little rip, and little sibling
  of the big rip} 

The first discussion of phantom fields in quantum cosmology was
presented in \cite{Dabrowski:2006dd}. The simplest case is a vanishing
potential and 
a vanishing cosmological constant. This is not a realistic model,
because it corresponds to stiff matter ($w=1$ in $p=w\rho$) and
$\rho<0$. Nevertheless, it is an instructive example because it can
be solved exactly. As mentioned above, this model has, in the phantom
case, only solutions for negative spatial curvature ${\mathcal
  K}=-1$ \cite{HCI18}. Classically, this universe collapses from
infinity, reaches a minimum for the scale factor (bounce) and
re-expands to infinity. At infinity, it exhibits a little rip abrupt
event. 

Solving the Wheeler--DeWitt equation (\ref{mini}) for this case, one
has to impose the boundary condition $\psi\stackrel{a \to
  0}{\longrightarrow}0$, because 
this region is classically forbidden and we thus demand the
wave function to go to zero there.\footnote{Although this is a natural
  condition and is standard in quantum mechanics, it is not always
  implemented in quantum cosmology. The wave functions following from
  the no-boundary (Hartle--Hawking) condition, for example, typically
  {\em increase} in classically forbidden regions \cite{annals91}.}
The solution is then a particular
Bessel function for $a$ and a plane wave for the phantom field. 
A more realistic model for the little rip is discussed in
\cite{Albarran:2016ewi}. As shown there, it is possible to apply the
DeWitt criterion and to find solutions of (\ref{mini}) for which the
wave functions vanish in the little rip region; in this sense, the
little rip can be avoided.

A real big-rip singularity is obtained for a phantom field with a
Liouville potential 
$V(\phi)=V_0\exp(-\lambda\kappa\phi)$ \cite{Dabrowski:2006dd}; 
classically, scale factor and
energy density diverge at finite time. A big bang singularity is not
present, as is true for all models with $w<-1$ and $\rho>0$
\cite{HCI18}. For this model, one can find {\em exact} wave-packet
solutions for the Wheeler--DeWitt equation. These solutions clearly
exhibit that wave packets necessarily {\em disperse} when approaching the region
of the classical big rip singularity. Following our second criterion
above, this can be taken as a signal of singularity avoidance: the
semiclassical approximation breaks down and the concept of an
expanding universe ceases to hold before the big rip region is
reached. This is an explicit example for the occurrence of quantum
gravitational behavior for large-size universes. Another example is
the occurrence of a quantum region when approaching the turning point
of a classically recollapsing universe \cite{KZ95}.

For negative cosmological constant and a phantom field, a model is
obtained in which the universe evolves between two big rips in a
finite time \cite{Dabrowski:2006dd}. It contracts from the first one,
reaches a minimum for the scale factor, and expands to the other
one. The phantom field has a potential containing a
$\cosh^2\phi$-term. For this model, the Wheeler--DeWitt equation can
be solved in the region near the big rips, and it is found again that
wave packets necessarily disperse, that is, one approaches the
timeless quantum region before reaching the classical singularities. 

Likewise, the quantum cosmology of models that induce a little rip and
a little sibling of the big rip 
have been analyzed in \cite{Albarran:2016ewi,Albarran:2015cda},
respectively. These analyses have been carried by invoking a perfect
fluid or a  
scalar field. In the case of matter described by a scalar field, the
Wheeler-DeWitt equation is the same as shown in (\ref{mini}).

Investigations are not restricted to FLRW models. It is possible to
discuss the presence of big rip and its quantum fate in the framework
of anisotropic models, notably Bianchi models \cite{KKP18}. 
The study of anisotropic models is an important step towards the
understanding of the general case (BKL conjecture). 

\subsubsection{Other phantom-induced singularities}

Big rip and little rip are intrinsic to the presence of phantom
fields. But phantoms allow the occurrence of singularities which
can also occur for ordinary fields.

Of particular interest is the situation where we have a
generalized Chaplygin gas effective equation of state,
\be
p=-\frac{A}{\rho^{\beta}},
\ee
with parameters $A$ and $\beta$.
Such an equation of state is of relevance when discussing dark energy
and dark matter. Phantom fields can induce big freeze (type~III) and big
d\'emarrage (type~II) singularities. 
In \cite{BouhmadiLopez:2009pu}, the Wheeler--DeWitt equation was
investigated for these cases\footnote{The quantum cosmology of a
  generalized Chaplygin gas was first carried in
  \cite{BouhmadiMoniz1}.} (and for ordinary scalar fields, too). It 
was shown that classes of solutions can be found that avoid the
singularities in the sense of the DeWitt criterion. This is also
possible for ordinary scalar fields, so phantoms do in this respect
not play a different role. 

The situation is somewhat different for 
type IV singularities \cite{Bouhmadi-Lopez:2013tua}. These
singularities, which are of a rather mild nature (for example,
geodesics are unaffected by it), are generically not avoided, only in
particular cases.

\subsubsection{Loop quantum cosmology}

Singularity avoidance of phantom (and other) fields has also been
discussed in loop quantum cosmology \cite{Bojowald,Calcagni}.
This is the application of loop quantum gravity to cosmology, with
loop quantum gravity being a variant of canonical quantum gravity,
distinguished from geometrodynamics by its different
use of variables. The central equation of loop quantum cosmology is
still of the form (\ref{wdw}), but it is claimed to have fundamentally
the form of a difference equation.  

The criteria for singularity avoidance are somewhat different from the
criteria used above (see e.g. the list in \cite{Calcagni},
p.~508). Much emphasis is taken on effective modifications of the Friedmann
equations. It follows that the matter density $\rho$ is replaced there
by
\be
\rho \longrightarrow \rho\left(1-\frac{\rho}{\rho_{\rm crit}}\right),
\ee
where $\rho_{\rm crit}$ denotes a critical density of the order of the
Planck density. This modification may introduce a bounce at small
scale factors and thus to an avoidance of the big bang singularity. 
A similar situation can occur for phantom-induced singularities. It
has, in fact, been shown that a big rip singularity is avoided
and replaced by a transition from an expanding to a contracting branch
\cite{SST06,SV11}. As in geometrodynamics, type IV singularities are
much less likely to be avoided.
Singularity avoidance for Bianchi spacetimes has been discussed in
\cite{WE18}.

%%%%%%%%%%%%%%%%%%%%%%%%%%%%%%%%
\subsection{Quantum cosmology of alternative theories of
  gravity}\label{sec:4.3} 

While so far the analysis of dark energy singularities has been mainly
focused on GR, it should be as well analyzed within alternative
theories of gravity where indeed these singularities can appear as we mentioned
previously. So, we consider here that these theories are classical
theories of gravity to be quantized, as they also predict
singularities. As there are many ways of extending GR, we will focus  
on the two main stream to define a modified theory of gravity: (i) a
metric a approach and (ii) a Palatini approach.

\subsubsection{Quantum cosmology in metric $f(R)$ gravity} 

Nowadays, $f(R)$ metric gravity is one of the best candidates not only
to describe the early inflationary era through the Starobinsky model, see
\cite{Starobinsky:1980te}, but
also to describe the late-time universe. 
The Starobinsky model was quantized back in the eighties by getting the
correct Wheeler--DeWitt equation for a homogeneous and isotropic
universe after introducing at the classical level a proper Lagrange
multiplier which takes into account the relation between the scale
factor and the scalar curvature \cite{Vilenkin}. It should be
highlighted that for a homogeneous and isotropic universe, the
Wheeler-DeWitt equation in this case has two degrees of freedom even
if the universe is empty \cite{Vilenkin}. Quantum cosmology for a more
general class of higher-derivative theories was discussed in
\cite{HL84,Horowitz}.  

The big rip singularity in the framework of $f(R)$ quantum
geometrodynamics and invoking the DeWitt criterion has been recently
analyzed in \cite{Alonso-Serrano:2018zpi}, where it was shown the
existence of solutions to the Wheeler--DeWitt equation fulfilling
this condition.  It is worthy to note that this equation is always
hyperbolic for any $f(R)$-cosmology, even if the classical model
mimics a phantom expansion.

\subsubsection{Quantum cosmology in Palatini EiBI gravity} 

We have already mentioned the EiBI theory. It has the bonus of
removing the big bang  singularity at the classical level and even
some phantom dark energy singularities, though not the big rip. This
has motivated the analysis carried in
\cite{Bouhmadi-Lopez:2016dcf,Albarran:2017swy,Bouhmadi-Lopez:2018tel,Albarran:2018mpg}. The
quantization of this theory has to be done with great care. In fact, a
thorough analysis of the classical Hamiltonian with constraints must
be carried out in order to get a  self-consistent modified Wheeler--DeWitt
equation. This new Wheeler--DeWitt  equation is derived with the use
of Dirac brackets. What should be highlighted in the quantization
of this theory is that the auxiliary scale factor, that is, the one
compatible with the Palatini connection, is the one that appear in the
Wheeler--DeWitt  equation rather than the standard scale factor. This
has important consequences when imposing boundary conditions close to
a singularity. 

In this kind of theories, it can be shown that the big rip singularity
present in the classical theory, and induced by a phantom perfect
fluid or a phantom scalar field, is expected to be removed when
quantum effects encoded on the modified Wheeler--DeWitt equation are
taken into account \cite{Bouhmadi-Lopez:2018tel,Albarran:2018mpg}. A
similar analysis has been carried for the little rip and little
sibling of the big rip reaching similar conclusions
\cite{Albarran:2017swy}.

%%%%%%%%%%%%%%%%%%%%%%%%%%%%%%%%%%%%%%%%%%%%%%%%%%%%%%%%%
\section{Discussion and perspectives}\label{sec:5}

So far, dark energy remains a mystery. Current observations allow the
possibility that it is caused by the presence of phantom fields or by
an alternative theory of gravity that mimicks their behavior. We have
reviewed here the properties of these fields and have discussed in
some detail the singularities caused by them in the classical theory
as well as their possible quantum avoidance. 

A possible modification of general relativity can arise from its
quantization or from its violation already at the classical
macroscopic level. Modifications from quantization are 
expected because it would not be natural to have a hybrid unified theory of
interactions in which one part (gravity) stays classical
\cite{OUP}. Quantum modifications are expected to lead to tiny
corrections in most situations and to become relevant when approaching
the region of classical singularities. The situation is different for
an alternative classical theory of gravity. Such a theory could
directly explain the dynamics of dark energy, without invoking any
additional fields, be them standard or phantom. But such a theory may
also predict the occurrence of singularities and would thus point to
the need of its quantization. So far, however, it is not known
whether important features of general relativity such as the initial
value problem or the existence of singularity theorems continue to
hold in alternative theories. Much work is thus needed to study
those theories in the classical and quantum realm \cite{Alonso-Serrano:2018zpi,Bouhmadi-Lopez:2018tel}. It is hard to
imagine that a decision on these issues can be reached without
observational input. It is therefore important to investigate whether
quantum gravitational effects such as the ones calculated
in \cite{BKK16,Bouhmadi-Lopez:2018bqw} can be observed and help us to reach such a decision.

%\noindent{\bf Note added in proof.}\\
\paragraph{Note added in proof.}
We have recently become aware of the paper ``Cosmological constraints from the Hubble diagram of quasars at high redshifts'' by Guido Risaliti and Elisabeta Lusso, in Nature Astronomy 3 (2019) 272–277, in which the authors present strong observational hints for dark energy caused by phantom fields ($w<−1.3$), using data of quasars and supernovae at large redshifts.

\begin{acknowledgements}
The work of MBL is supported by the Basque Foundation of Science
IKERBASQUE. She also wishes to acknowledge the partial support from
the Basque government Grant No. IT956-16 (Spain) and and FONDOS FEDER
under grant FIS2017- 
85076-P (MINECO/AEI/FEDER, UE).
PMM acknowledges financial support from the project FIS2016-78859-P (AEI/FEDER, UE). 
This article is based upon work from COST Action CA15117 (CANTATA),
supported by COST (European Cooperation in Science and Technology)
www.cost.eu. 
\end{acknowledgements}

%%%%%%%%%%%%%%%%%%%%%%%%%%%%%%%%
%%%%%%%%%%%%%%%%%%%%%%%%%%%%%%%%

\end{document}